# Laser agitates probability flow in atoms to form alternating current and its peak-dip phenomenon

Huai-Yang Cui

Department of Physics, Beihang University

Xueyuan Road 37, Beijing, 100191, China.

E-mail:hycui@buaa.edu.cn

Abstract: By using trajectory-based approaches to quantum transition, it is found that laser can agitate the probability flow in atoms to form alternating current with the frequency of the laser. The detailed physical process of quantum transition is investigated, during which the alternating current in atomic probability flow becomes a key role connecting the external electromagnetic wave with the evolution of the quantum states in atoms. Computer was employed to simulate the physical process. The atomic alternating current may have the peak-dip phenomenon.

The trajectory-based approaches to quantum mechanics have recently attracted widespread attention as a powerful tool for investigating fundamental questions in quantum mechanics[1-5], and have generated excitement for enhancing precise single photon measurement[6,7]. These approaches are often referred to as de Broglie-Bohm theory[8-10], or hidden variable theory[11-13]. In this paper we apply the de Broglie-Bohm theory to quantum transition that happens in a short time none longer than $10^{-9}$ sec.[14], ones have an impression that the transition means sudden shift with a negligible process. In fact, there is an un-negligible physical process for quantum transition when we know how a laser to agitate the probability flows in atoms.

We begin with the coherent length of de Broglie matter wave, as shown in Fig.1, if the coherent length of the electron in hydrogen atom is long enough, it will wind around the time axis, the matter wave in the first circle will overlap and interfere with itself tail at the same location. The overlapping number $N$ depends on the coherent length; if $N=\infty$, the matter wave has the interference like the Fabry-Perot interference in optics.

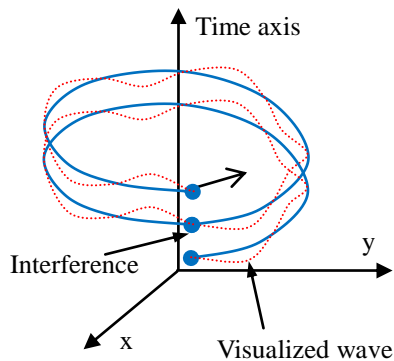

Fig.1 The matter wave winds around the time axis, overlap and interfere with itself in a form of the Fabry-Perot interference.

de Broglie matter wave in the first circle is given by

$$w = \exp(\frac{ipl}{\hbar} - \frac{iEt}{\hbar}) \tag{1}$$

where $l$ is the length of the path along the orbit from the initial point $(r = r_0, \theta = 0)$. The overlapped matter wave is given by

$$\psi = w + wae^{i\delta} + wa^2 e^{i2\delta} + \ldots = \frac{w}{1-a\exp(i\delta)} \tag{2}$$

where $\delta$ is the phase shift after one circle retardation for the matter wave; $a$ denotes attenuation per cycle. Obviously, Only the Bohr's orbits can survive in the denominator of Eq.(2), they satisfy

$$\delta = \frac{1}{\hbar} \oint_L p\, dl = 2\pi n; \quad n = 1, 2, 3, \ldots \tag{3}$$

If a laser comes in, the electron transits from the state 1 to the state 2, its energy and momentum change from $(E_1, P_1)$ to $(E_2, P_2)$ respectively, its matter wave in the first circle also changes

$$w_1 = \exp(\frac{ip_1 l}{\hbar} - \frac{iE_1 t}{\hbar}) = \exp(\frac{ip_2 l}{\hbar} - \frac{iE_2 t}{\hbar}) \exp\left(\frac{i(p_1 - p_2)l}{\hbar} - \frac{i(E_1 - E_2)t}{\hbar}\right) \tag{4}$$

Introducing the notation

$$\delta = \frac{(p_1 - p_2)l}{\hbar} - \frac{(E_1 - E_2)t}{\hbar} \tag{5}$$

suppose the matter wave in the state 2 smoothly inherits the state 1 through

$$w_2 \approx w_1; \quad w_2 = \exp(\frac{ip_2 l}{\hbar} - \frac{iE_2 t}{\hbar}) \exp(i\delta) \tag{6}$$

then the electron has an extra phase shift $\delta$ in the orbit of the state 2. When the Fabry-Perot interference occurs in the state 2, circle by circle, the overlapped wave is given by

$$\psi = \exp(\frac{ip_2 l}{\hbar} - \frac{iE_2 t}{\hbar}) \frac{1}{1 - a\exp(i\delta)} \tag{7}$$

$$\frac{1}{\hbar} \oint_L p_2\, dl = 2\pi n_2; \quad n_2 = 1, 2, 3, \ldots \tag{8}$$

Substituting it into probability flow formula, we obtain

$$\begin{aligned}
j &= \frac{\hbar}{2im_e}\left(\psi^* \frac{\partial \psi}{\partial l} - \psi \frac{\partial \psi^*}{\partial l}\right) \\
&= \frac{[p_2 - (p_1 - p_2)/2]/m_e}{1 + a^2 - 2a\cdot\cos(\delta)} + \frac{[(1-a^2)(p_1 - p_2)/2]/m_e}{[1 + a^2 - 2a\cdot\cos(\delta)]^2} \\
&= \frac{[p_2 - (p_1 - p_2)/2]/m_e}{1 + a^2 - 2a\cdot\cos[(p_1 - p_2)l/\hbar - (E_1 - E_2)t/\hbar]} \\
&\quad + \frac{[(1-a^2)(p_1 - p_2)/2]/m_e}{[1 + a^2 - 2a\cdot\cos[(p_1 - p_2)l/\hbar - (E_1 - E_2)t/\hbar]]^2}
\end{aligned} \tag{9}$$

It is alternating current as shown in Fig.2 with the frequency $\omega = |E_1 - E_2|/\hbar$, the atom responds positively to the external electromagnetic wave. This is just what we expect for the radiation mechanism: if the electron gains the energy $|E_1 - E_2|$, it will move in a way that it seems in an alternating current with the frequency $\omega = |E_1 - E_2|/\hbar$ and digests the electromagnetic wave energy with the frequency $\omega = |E_1 - E_2|/\hbar$.

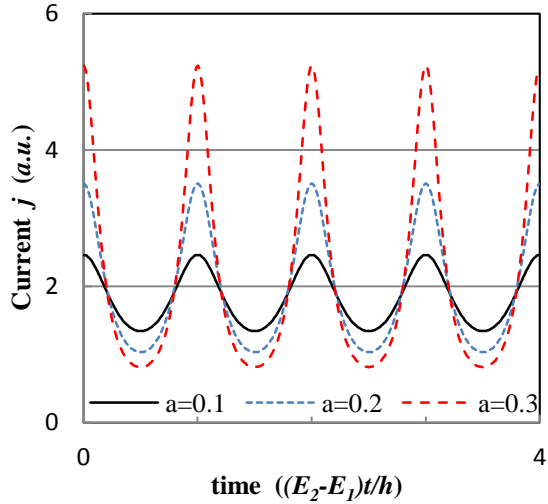

Fig.2 Alternating current of probability flow in atoms at the location $l = 0$.

To pay attention to the factor $(p_1 - p_2)l/\hbar$ in the alternating current formula, only it is a multiple of $2\pi$ after one orbital cycle, it can vanish from the denominator of Eq.(9), thus the alternating current can achieve in macroscopic scale. This factor in integral form is given by

$$\frac{1}{\hbar}\oint_L (p_1 - p_2)dl = 2\pi k; \quad k = 0, \pm 1, \pm 2, \ldots \quad (10)$$

As shown in Fig.3, the atomic AC of the probability flow is visualized by the gray in the ring. When $k = \pm 1$, the atomic AC seems to be a dipole rotating in the orbit, see Fig.3(a); when $k = \pm 2$, the atomic AC seems to be a quadrapole rotating in the orbit, see Fig.3(b). According to classical electrodynamics, dipole model has the most strong ability to absorb or emit electromagnetic wave, therefore the most effective transition takes place at $k = \pm 1$, so that in practice the **selection rule** for atomic radiation is $k = \pm 1$. Especially, the radiation of $k = 0$ transition is forbidden absolutely because of non pole in the orbit, its probability flow $j(t)$ has the same value in the orbit at a time, but AC part of $j(t)$ is oscillating on the time axis (the charge twinkles).

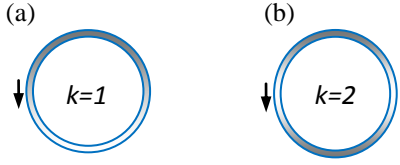

(a) Dipole model for atomic AC.
(b) Quadrapole model for atomic AC.

Fig. 3 Atomic AC of the probability flow is visualized by the gray in the ring.

The reverse steps in the above formulism are suitable for the atomic emission to the electromagnetic radiation. The mechanism can also be applied to LED.

How can the election stays in the Bohr's orbit to avoid orbital collapse while its speed changes due to the alternating current? Actually, it has its stable mechanism. Coherent width is a counter-part to coherent length, rarely mentioned in the textbooks. We consider a Bohr's orbit with the radius $r_0$ and the speed $v_0$ in the ground state, the matter wave with a width is supposed in the first circle to be

$$\phi = \exp(\frac{im_e v_0 r\theta}{\hbar} - \frac{iEt}{\hbar}) \tag{11}$$

Separating it into two parts, we focus on the extra phase shift in the orbit

$$\phi = \exp(\frac{im_e v_0 r_0 \theta}{\hbar} - \frac{iEt}{\hbar})\exp(\frac{im_e v_0 (r-r_0)\theta}{\hbar}) \tag{12}$$

Because the wave winds around the time axis and overlaps itself, following Eq.(2), the overlapped wave is given by

$$\psi = w + wae^{i\delta} + wa^2 e^{i2\delta} + ... + wa^N e^{iN\delta} = w\frac{1-a\exp(iN\delta)}{1-a\exp(i\delta)} \tag{13}$$

$$w = \exp(\frac{im_e v_0 r_0 \theta}{\hbar} - \frac{iEt}{\hbar}) \tag{14}$$

$$\delta = \frac{2\pi m_e v_0 (r-r_0)}{\hbar} \tag{15}$$

Where $N$ is the overlapping number. For $N = \infty$, it is easy to find the quantum force in the radial direction

$$f_r = \frac{d}{dr}\left(\frac{1}{2m_e}P_r^2\right) = \frac{1}{2m_e}\frac{d}{dr}\left(-i\hbar\frac{\partial \ln \psi}{\partial r}\right)^2 \tag{16}$$

where $P_r$ is the momentum carried by the matter wave in radial direction. As shown in Fig.4, the quantum force $f_r$ definitely locks the Bohr's orbit and traps the electron in radial direction, even when the alternating current occurs in the orbit. Computer was employed to simulate the stable mechanism due to the quantum force, as shown in Fig.5, where 100 electrons emit respectively at the initial speed uniformly from $v = 0.5v_0$ to $v = 1.5v_0$ at $r_0$ in the direction of $\theta$, after 1 period, their final locations falls into the narrow range about the radius $r_0$, The location density is in a stable distribution due to the quantum force Eq.(16) in radial direction.

In the point of the author's view[15], the quantum force $f_r$ occupies the freedom that belongs to the quantum hidden variable, does not conflict with the freedom of the Coulomb's action.

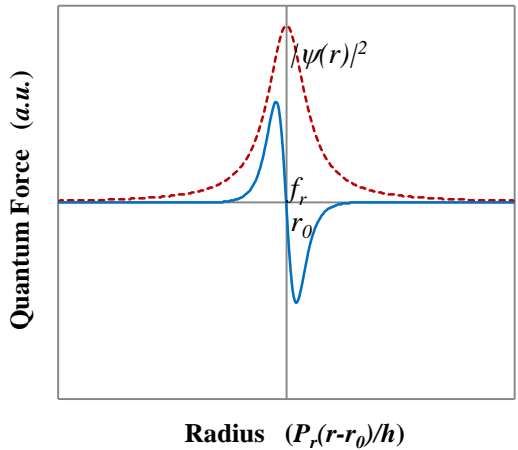

Fig.4 the quantum force $f_r$ definitely locks the Bohr's orbit and traps the electron in radial direction.

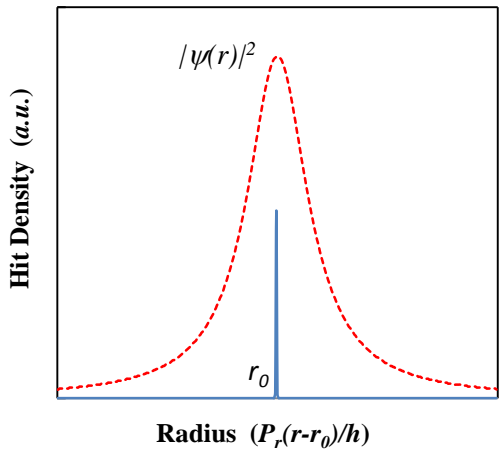

Fig.5 Computer simulation: 100 electrons emit respectively at the initial speed from $v = 0.5v_0$ to $v = 1.5v_0$ at $r_0$ in the direction of $\theta$, after 1 period, their final location distribution is stable.

The overlapping number $N$ depends on the coherent length (see Fig.1), which can be experimentally determined by measuring the width of light spectral line. Since the coherent length is finite, $N = \infty$ seems irrational. As shown in Fig.6, $N = 20$ case leads to the matter wave in radial direction to split into two striking maxima about the Bohr's radius $r_0$, with a feature of the peak-dip (Lamb dip[16]) at the $r_0$. $N = 10$ case and $N = 5$ case all show the peak-dips.

Nevertheless, the peak-dip means that the single Bohr's orbit splits into double secondary orbits that could accommodate one spin-up electron and another spin-down electron; one orbit contains a pair electrons, like helium in the ground state. Not only the hydrogen atom, the famous double lines in sodium light spectrum experimentally demonstrate spin-up electron & spin-down electron sharing one orbit in alkali metal atoms.

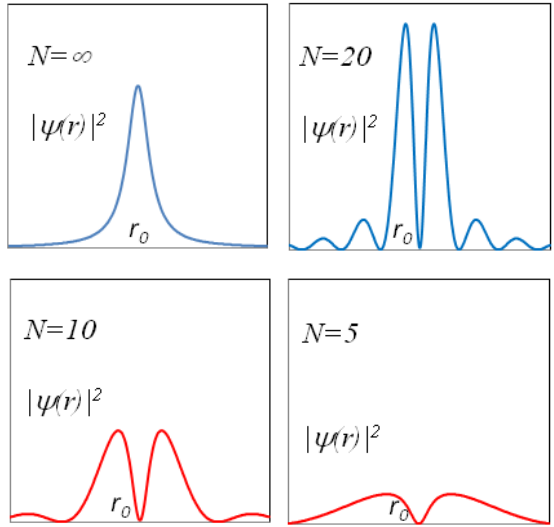

Fig.6  $N = 20$  case leads the matter wave in radial direction to split into striking two maxima about the Bohr's radius $r_0$.

As the related parameters vary, one Bohr's orbit may split into other multi-orbits structure, but double-splitting is the most striking one in the nature. Because of $E = -e^2/(8\pi\varepsilon r)$ for hydrogen atom, it is easy to transform $\delta(r) \to \delta(E)$, so the peak-dip of $\delta(r)$ will transform into the peak-dip of the light spectrum according to $\delta(E)$. Fig.7 shows the recoil-induced spectral doubling of the $CH_4$ saturated absorption peaks at 3.39μm with the three peak-dips[17], here it is explained as the orbit-splitting in terms of trajectory-based approaches. The same explanation gives to the fine structure light spectrum of $NO_2$ in the rotation transition $J'' = 0 \to J' = 0$ in Fig.8 with the several peak-dips[18].

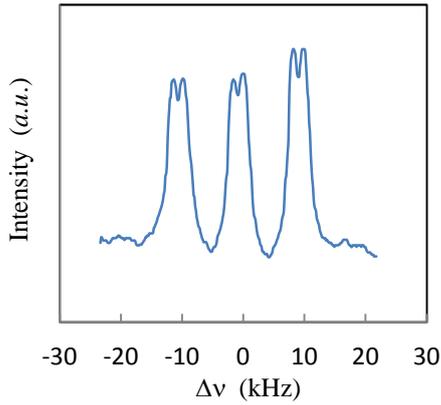

Fig.7 The fine structure light spectrum of $CH_4$ at 3.39μm with 3 peak-dips. Copyright: Phys. Rev. Lett. [17]

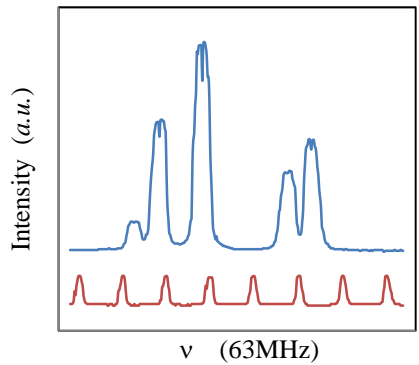

Fig.8 The fine structure light spectrum of $NO_2$ in the rotation transition $J''=0 \to J'=0$. Copyright: Chem. Phys. [18]

It is easy to experimentally verify the atomic alternating current **in an indirect way**, because the phase shift $\delta(t)$ shares the same mathematical expression with other kinds of phase shifts, that is

$$\psi = w \frac{1 - a\exp(iN\delta)}{1 - a\exp(i\delta)} \begin{cases} \delta(r); & \text{orbit splitting, spin up and down} \\ \delta(L); & \text{Lamb dip for lase cavity length L(?)} \\ \delta(E); & \text{peak-dip in light spectrum} \\ \delta(t); & \text{peak-dip on time axis, Atomic AC} \end{cases} \quad (17)$$

When $N$ is finite, the peak-dip on the light spectrum will transform into the peak-dip on the atomic AC, sharing the same topology. The peak-dip on the atomic AC causes the atom to emit the electromagnetic wave that contains multi-frequencies: $\omega, 2\omega, 3\omega$, etc, because the peak-dip approximately doubles the frequency, a peak-dip adds a crest and a trough into the original period. In 1961, a ruby laser 694nm illuminated into a quartz sample and output a light 347nm [19], here it is explained as the existence of peak-dips on its atomic AC.

Conclusions: By using trajectory-based approaches to quantum transition, it is found that laser can agitate the probability flow in atoms to form alternating current with the frequency of the laser. The detailed physical process of quantum transition is investigated, during which the alternating current in atomic probability flow becomes a key role connecting the external electromagnetic wave with the evolution of the quantum states in atoms. Computer was employed to simulate the physical process. The atomic alternating current may have the peak-dip phenomenon.